\documentclass[12pt]{article}

\usepackage{axodraw}
\usepackage{epsf}
\usepackage{rotating}

\newcommand{\beq}{\begin{equation}}
\newcommand{\eeq}{\end{equation}}
\newcommand{\bea}{\begin{eqnarray}}
\newcommand{\eea}{\end{eqnarray}}
\newcommand{\bed}{\begin{displaymath}}
\newcommand{\eed}{\end{displaymath}}

\newcommand{\tgb}{{\rm tg}\beta}

\newcommand{\gl}{\tilde g}

\newcommand{\lsim}{\raisebox{-0.13cm}{~\shortstack{$<$ \\[-0.07cm] $\sim$}}~}

\catcode`\@=11

\def\@citex[#1]#2{\if@filesw\immediate\write\@auxout{\string\citation{#2}}\fi
  \def\@citea{}\@cite{\@for\@citeb:=#2\do
    {\@citea\def\@citea{,\penalty\@m}\@ifundefined
       {b@\@citeb}{{\bf ?}\@warning
       {Citation `\@citeb' on page \thepage \space undefined}}%
\hbox{\csname b@\@citeb\endcsname}}}{#1}}

\def\citer{\@ifnextchar [{\@tempswatrue\@citexr}{\@tempswafalse\@citexr[]}}
\def\@citexr[#1]#2{\if@filesw\immediate\write\@auxout{\string\citation{#2}}\fi
  \def\@citea{}\@cite{\@for\@citeb:=#2\do
    {\@citea\def\@citea{--\penalty\@m}\@ifundefined
       {b@\@citeb}{{\bf ?}\@warning
       {Citation `\@citeb' on page \thepage \space undefined}}%
\hbox{\csname b@\@citeb\endcsname}}}{#1}}
\catcode`\@=12

\begin{document}

\renewcommand{\thefootnote}{\fnsymbol{footnote}}
\setcounter{page}{0}

\begin{titlepage}

\vskip-1.0cm

\begin{flushright}
PSI--PR--08--15 \\
LAPTH--1295/08 \\
KA--TP--33--2008 \\
SFB/CPP--08--106
\end{flushright}

\begin{center}
{\large \sc MSSM Higgs Boson Production via Gluon Fusion:} \\[0.5cm]
{\large \sc The Large Gluino Mass Limit}\\
\end{center}

\vskip 1.cm
\begin{center}
{\sc Margarete M\"uhlleitner$^1$, Heidi Rzehak$^2$ and Michael Spira$^3$}

\vskip 0.8cm

\begin{small} 
{\it \small
$^1$ LAPTH, Universit\'e de Savoie, CNRS, BP110, F--74941
Annecy-le-Vieux Cedex, France \\
$^2$ ITP, Universit\"at Karlsruhe, D--76128 Karlsruhe, Germany \\
$^3$ Paul Scherrer Institut, CH--5232 Villigen PSI, Switzerland}
\end{small}
\end{center}

\vskip 2cm

\begin{abstract}
\noindent
Scalar MSSM Higgs boson production via gluon fusion $gg\to h,H$ is
mediated by heavy quark and squark loops. The higher order QCD
corrections to these processes turn out to be large. The full
supersymmetric QCD corrections have been calculated recently. In the limit
of large SUSY masses a conceptual problem appears, i.e.~the proper
treatment of the large gluino mass limit. In this work we will describe
the consistent decoupling of heavy gluino effects and derive the
effective Lagrangian for decoupled gluinos.
\end{abstract}

\end{titlepage}

\renewcommand{\thefootnote}{\arabic{footnote}}

\setcounter{footnote}{0}

\section{Introduction}
Higgs boson \cite{hi64} searches belong to the primary motivations for
present and future colliders.  In the MSSM two isospin Higgs doublets
are introduced for the masses of up- and down-type fermions
\cite{twoiso}. After electroweak symmetry breaking five states are left
as elementary Higgs particles, two CP-even neutral (scalar) particles
$h,H$, one CP-odd neutral (pseudoscalar) particle $A$ and two charged
bosons $H^\pm$. At leading order (LO) the MSSM Higgs sector is fixed by
two independent input parameters which are usually chosen as the
pseudoscalar Higgs mass $M_A$ and $\tgb=v_2/v_1$, the ratio of the two
vacuum expectation values. Being lighter than the $Z$ boson mass at LO,
the one-loop and dominant two-loop corrections shift the upper bound of
the light scalar Higgs mass to $M_h\lsim 140$ GeV \cite{mssmrad}.  The
couplings of the various neutral Higgs bosons to fermions and gauge
bosons, normalized to the SM Higgs couplings, are listed in
Table~\ref{tb:hcoup}. The angle $\alpha$ denotes the mixing angle
of the scalar Higgs bosons $h,H$.  An important property of the bottom
Yukawa couplings is their enhancement for large values of $\tgb$, while
the top Yukawa couplings are suppressed for large $\tgb$
\cite{schladming}. Thus, the top Yukawa couplings play a dominant role
at small and moderate values of $\tgb$.
\begin{table}[hbt]
\renewcommand{\arraystretch}{1.5}
\begin{center}
\begin{tabular}{|lc||ccc|} \hline
\multicolumn{2}{|c||}{$\phi$} & $g^\phi_u$ & $g^\phi_d$ &  $g^\phi_V$ \\
\hline \hline
SM~ & $H$ & 1 & 1 & 1 \\ \hline
MSSM~ & $h$ & $\cos\alpha/\sin\beta$ & $-\sin\alpha/\cos\beta$ &
$\sin(\beta-\alpha)$ \\ & $H$ & $\sin\alpha/\sin\beta$ &
$\cos\alpha/\cos\beta$ & $\cos(\beta-\alpha)$ \\
& $A$ & $ 1/\tgb$ & $\tgb$ & 0 \\ \hline
\end{tabular}
\renewcommand{\arraystretch}{1.2}
\caption[]{\label{tb:hcoup} \it Higgs couplings in the MSSM to fermions
and gauge bosons [$V=W,Z$] relative to the SM couplings.}
\end{center}
\end{table}

Usually the scalar superpartners $\tilde f_{L,R}$ of the left- and
right-handed fermion components mix with each other. However, in this
work we will neglect mixing effects. Thus the masses of the sfermion
states $\tilde f_{L,R}$ are simply given by
\begin{equation}
M_{\tilde f_{L,R}}^2 = m_f^2 + M_{L,R}^2
\end{equation}
where $m_f$ denotes the fermion mass and $M_{L/R}$ the soft
supersymmetry-breaking sfermion mass parameters.  The neutral scalar
Higgs couplings [${\cal H}=h,H$] to non-mixing sfermions read
\cite{DSUSY}
\begin{eqnarray}
g_{\tilde f_L \tilde f_L}^{\cal H} & = & g_{\tilde f_R \tilde f_R}^{\cal
H} = m_f^2 g_f^{\cal H} \nonumber \\
g_{\tilde f_L \tilde f_R}^{\cal H} & = & 0 \nonumber \\
g_{\tilde f_i \tilde f_j}^A & = & 0
\label{eq:hsfcouprl}
\end{eqnarray}
with the couplings $g_f^{\cal H}$ listed in Table \ref{tb:hcoup}. $D$
terms have been neglected in these expressions. It is important to note
that supersymmetry relates the diagonal couplings to the corresponding
fermion Yukawa coupling involving the fermion mass $m_f$.

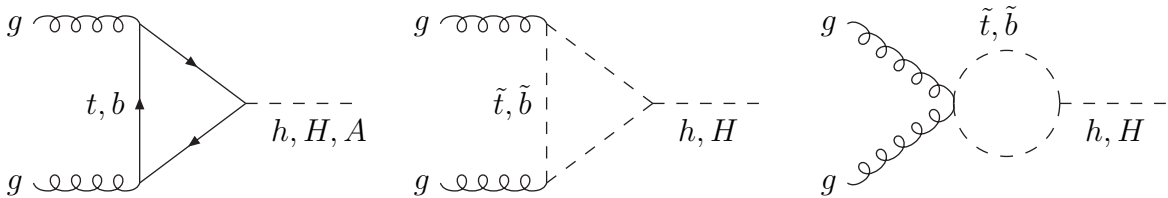
\begin{figure}[htb]
\begin{center}
\begin{picture}(130,90)(20,0)
\Gluon(10,20)(50,20){-3}{4}
\Gluon(10,80)(50,80){3}{4}
\ArrowLine(50,20)(50,80)
\ArrowLine(50,80)(90,50)
\ArrowLine(90,50)(50,20)
\DashLine(90,50)(130,50){5}
\put(0,18){$g$}
\put(0,78){$g$}
\put(30,46){$t,b$}
\put(100,36){$h,H,A$}
\end{picture}
\begin{picture}(130,90)(0,0)
\Gluon(10,20)(50,20){-3}{4}
\Gluon(10,80)(50,80){3}{4}
\DashLine(50,20)(50,80){5}
\DashLine(50,80)(90,50){5}
\DashLine(90,50)(50,20){5}
\DashLine(90,50)(130,50){5}
\put(0,18){$g$}
\put(0,78){$g$}
\put(30,46){$\tilde t,\tilde b$}
\put(100,36){$h,H$}
\end{picture}
\begin{picture}(130,90)(-20,0)
\Gluon(10,20)(50,50){-3}{5}
\Gluon(10,80)(50,50){3}{5}
\DashCArc(70,50)(20,180,360){5}
\DashCArc(70,50)(20,0,180){5}
\DashLine(90,50)(130,50){5}
\put(0,18){$g$}
\put(0,78){$g$}
\put(60,76){$\tilde t,\tilde b$}
\put(100,36){$h,H$}
\end{picture}  \\
\caption{\label{fg:lodiahgg} \it MSSM Higgs boson production via gluon
fusion mediated by top- and bottom quark as well as stop and sbottom
loops at leading order.}
\end{center}
\end{figure}

At hadron colliders as the Tevatron and LHC neutral Higgs bosons are
copiously produced by the gluon fusion processes $gg\to h/H/A$, which
are mediated by top and bottom quark loops as well as stop and sbottom
loops for the scalar Higgs bosons $h,H$ in the MSSM (see
Fig.~\ref{fg:lodiahgg}) \cite{cxn}.

The pure QCD corrections to the (s)top and (s)bottom quark loops are
known including the full Higgs and (s)quark mass dependences
\cite{gghnlo}.  They increase the cross sections by up to about 100\%.
The limit of very heavy top quarks and squarks provides an approximation
within $\sim 20-30\%$ for $\tgb\lsim 5$ \cite{limit}. In this limit the
next-to-leading order (NLO) QCD corrections have been calculated
\cite{gghnlolim} and later the next-to-next-to-leading order (NNLO) QCD
corrections \cite{gghnnlo}. The NNLO corrections lead to a further
moderate increase of the cross section by $\sim 20-30\%$, so that the
dominant part are the NLO contributions. An estimate of the
next-to-next-to-next-to-leading order (NNNLO) effects has been obtained
\cite{gghn3lo} indicating improved perturbative convergence.  Moreover,
the full SUSY--QCD corrections have been derived for heavy SUSY particle
masses \citer{gghnlosqcd,gghnlosqcd2} and recently including the full
mass dependence \cite{gghnlofull}. Ref.~\cite{gghnlosqcd} addresses the
limit of large gluino masses for the scalar Higgs couplings to gluons
for degenerate squark masses, i.e.~without mixing, as a special limit of
their final result. The result develops a logarithmic singularity for
large gluino masses which seems to contradict the decoupling of the
gluino contributions according to the Appelquist--Carazzone theorem
\cite{appcar}. In the pseudoscalar Higgs case this logarithmic
divergence for large gluino masses does not appear \cite{gganlosqcd}.
This work describes the resolution of this problem and a consistent
derivation of the effective Lagrangian after decoupling the gluino
contributions.

The paper is organized as follows. In Section 2 we derive the effective
Lagrangian in the limit of heavy quark, squark and gluino masses, where
the gluinos are much heavier than the quarks and squarks in addition.
Section 3 summarizes and concludes.

\section{Decoupling of the Gluinos}
\begin{figure}[htb]
\begin{center}
\begin{picture}(130,100)(0,0)
\DashLine(0,50)(50,50){5}
\ArrowLine(50,50)(100,100)
\ArrowLine(100,0)(50,50)
\put(5,55){${\cal H}$}
\put(105,98){$Q$}
\put(105,-2){$\bar Q$}
\Line(85,50)(60,50)
\ArrowLine(62,50)(60,50)
\put(95,48){$-i\lambda_Q$}
\put(50,-10){$(a)$}
\end{picture}
\begin{picture}(130,100)(-40,0)
\DashLine(0,50)(50,50){5}
\DashLine(50,50)(100,100){5}
\DashLine(100,0)(50,50){5}
\put(5,55){${\cal H}$}
\put(105,98){$\tilde Q$}
\put(105,-2){$\bar{\tilde Q}$}
\Line(85,50)(60,50)
\ArrowLine(62,50)(60,50)
\put(95,48){$-i\lambda_{\tilde Q}$}
\put(50,-10){$(b)$}
\end{picture}  \\
\begin{picture}(130,150)(0,-10)
\DashLine(0,50)(50,50){5}
\ArrowLine(75,75)(100,100)
\ArrowLine(100,0)(75,25)
\Line(75,75)(75,25)
\Gluon(75,25)(75,75){-3}{4}
\DashLine(50,50)(75,75){5}
\DashLine(75,25)(50,50){5}
\put(5,55){${\cal H}$}
\put(50,65){$\tilde Q$}
\put(80,48){$\tilde g$}
\put(105,98){$Q$}
\put(105,-2){$\bar Q$}
\put(50,-10){$(c)$}
\end{picture}
\begin{picture}(130,150)(-40,-10)
\DashLine(0,50)(50,50){5}
\DashLine(75,75)(100,100){5}
\DashLine(100,0)(75,25){5}
\Line(75,75)(75,25)
\Gluon(75,25)(75,75){-3}{4}
\ArrowLine(50,50)(75,75)
\ArrowLine(75,25)(50,50)
\put(5,55){${\cal H}$}
\put(50,65){$Q$}
\put(80,48){$\tilde g$}
\put(105,98){$\tilde Q$}
\put(105,-2){$\bar{\tilde Q}$}
\put(50,-10){$(d)$}
\end{picture}  \\
\caption{\label{fg:dia} \it Scalar MSSM Higgs boson couplings to heavy
quarks $Q$ and squarks $\tilde Q$: (a) ${\cal H}Q\bar Q$ coupling at LO,
(b) ${\cal H}\tilde Q\bar{\tilde Q}$ coupling at LO, (c) gluino
contribution to the ${\cal H}Q\bar Q$ coupling at NLO and (d) gluino
contribution to the ${\cal H}\tilde Q\bar{\tilde Q}$ coupling at NLO.
[${\cal H}=h,H$]}
\end{center}
\end{figure}
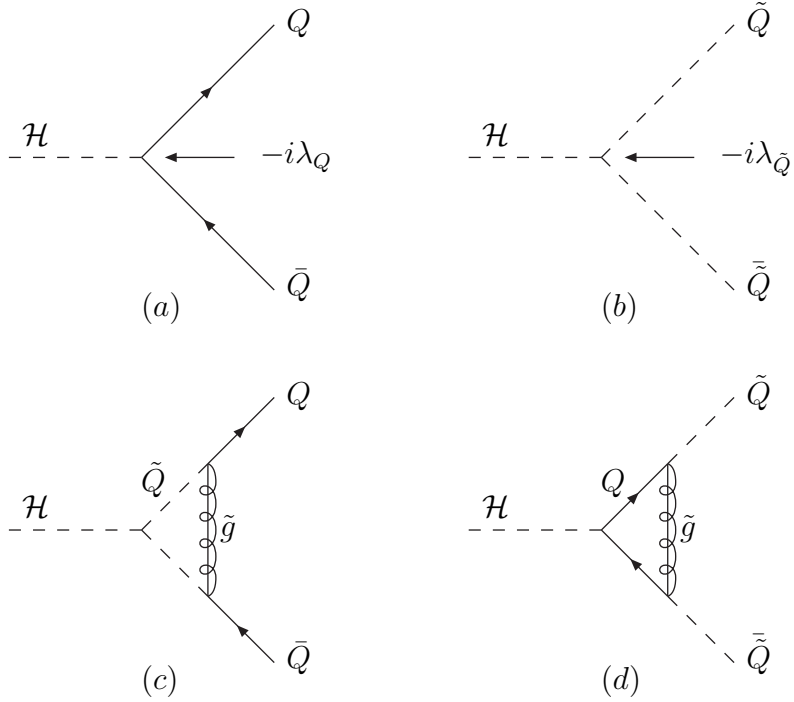

For the derivation of the effective Lagrangian for the scalar Higgs
couplings to gluons we will analyze the relation between the quark
Yukawa coupling $\lambda_Q$ and the Higgs coupling to squarks
$\lambda_{\tilde Q}$ in the limit of large gluino masses in detail. To
set up our notation we will define these couplings at leading order in
the case of vanishing squark mixing as
\begin{eqnarray}
\lambda_Q & = & g_Q^{\cal H} \frac{m_Q}{v} \nonumber \\ \lambda_{\tilde Q} &
= & 2\frac{g^{\cal H}_{\tilde f_L\tilde f_L}}{v} = 2\frac{g^{\cal H}_{\tilde
f_R\tilde f_R}}{v} = 2g_Q^{\cal H} \frac{m_Q^2}{v} = \kappa \lambda^2_Q
\nonumber \\
\kappa & = & 2\frac{v}{g_Q^{\cal H}}
\label{eq:cplg}
\end{eqnarray}
where $v=1/\sqrt{\sqrt{2}G_F} \approx 246$ GeV denotes the vacuum
expectation value of the Higgs sector which is related to the Fermi
constant $G_F$. These couplings mediate the scalar Higgs decays into
heavy quark pairs ${\cal H}\to Q\bar Q$ and into squark pairs ${\cal
H}\to \tilde Q\bar{\tilde Q}$ (see Fig.~\ref{fg:dia}a,b). In the
following we will derive the modified relation between these couplings
for scales {\it below} the gluino mass $M_{\tilde g}$. This will proceed
along the following lines: {\it (i)} We will start with the unbroken
relation between the running $\overline{MS}$ couplings of
Eq.~(\ref{eq:cplg}) for scales {\it above} the gluino mass and the
corresponding renormalization group equations. {\it (ii)} If the scales
decrease {\it below} the gluino mass the gluino decouples from the
renormalization group equations. This decoupling leads to modified
renormalization group equations which differ for the two couplings
$\lambda_{\tilde Q}$ and $\kappa \lambda^2_Q$. This implies that the two
couplings deviate for scales {\it below} the gluino mass, while they are
identical for scales {\it above} the gluino mass. {\it (iii)} The proper
matching at the gluino mass scale yields a finite threshold contribution
for the evolution from the gluino mass to smaller scales, while the
logarithmic structure of the matching relation is given by the solution
of the renormalization group equations {\it below} the gluino mass
scale. We will determine these ingredients in detail in this letter. The
matching relations can be obtained from the gluino contributions for
heavy gluino masses in the limit of vanishing external momentum
transfers \cite{colwilzee} as we will discuss in the next sections in
detail.

\subsection{\boldmath{$\phi\to Q\bar Q$}}
The gluino contribution to the Higgs vertex with heavy quark pairs is
depicted in Fig.~\ref{fg:dia}c. We use dimensional regularization in
$n=4-2\epsilon$ dimensions for the evaluation of the one-loop
contributions. Since all virtual particles are massive, there are no
infrared divergences. Although dimensional regularization requires the
introduction of anomalous counter terms in general to restore the
supersymmetric relations between corresponding couplings and masses,
these gluino contributions are free of these terms. The result of the
vertex contribution in the heavy gluino mass limit vanishes,
\begin{equation}
Z_1-1 \to 0
\end{equation}
\begin{figure}[htb]
\begin{center}
\begin{picture}(130,100)(30,-10)
\ArrowLine(0,50)(50,50)
\ArrowLine(100,50)(150,50)
\DashCArc(75,50)(25,180,360){5}
\GlueArc(75,50)(25,0,180){3}{9}
\CArc(75,50)(25,0,180)
\put(5,55){$Q$}
\put(135,55){$Q$}
\put(70,85){$\tilde g$}
\put(70,10){$\tilde Q$}
\put(67,-10){$(a)$}
\end{picture}
\begin{picture}(130,100)(-30,-10)
\DashLine(0,50)(50,50){5}
\DashLine(100,50)(150,50){5}
\ArrowArc(75,50)(25,180,360)
\GlueArc(75,50)(25,0,180){3}{9}
\CArc(75,50)(25,0,180)
\put(5,55){$\tilde Q$}
\put(135,55){$\tilde Q$}
\put(70,85){$\tilde g$}
\put(70,10){$Q$}
\put(67,-10){$(b)$}
\end{picture}  \\
\caption{\label{fg:self} \it Gluino contributions to the self-energies of
(a) quarks $Q$ and (b) squarks $\tilde Q$ at NLO.}
\end{center}
\end{figure}
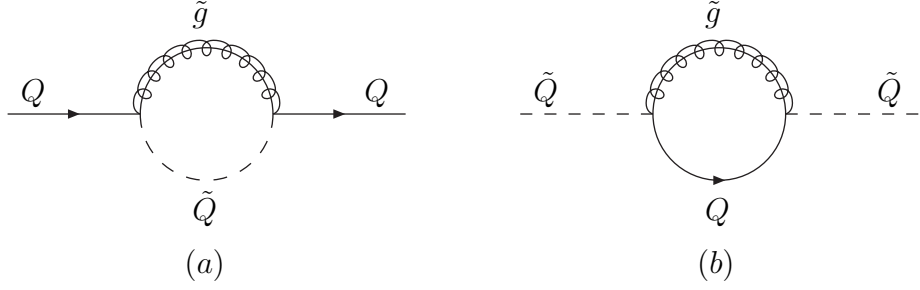
The full correction to the bare Yukawa coupling requires the addition of
the wave function renormalization constant $Z_2$ which can be derived
from the derivative of the corresponding self-energy diagram shown in
Fig.~\ref{fg:self}a. In the limit of large gluino masses the gluino
contribution is found to be [$C_F=4/3$]
\begin{equation}
Z_2-1 \to C_F \frac{\alpha_s}{\pi} \Gamma(1+\epsilon) \left(\frac{4\pi
\mu^2}{M_{\gl}^2}\right)^\epsilon \left\{ -\frac{1}{4\epsilon} -
\frac{3}{8} \right\}
\end{equation}
The gluino contributions to the bare quark Yukawa coupling can now be
derived as
\begin{equation}
\frac{\Delta \lambda_Q}{\lambda_Q} = Z_1Z_2-1 \to C_F \frac{\alpha_s}{\pi}
\Gamma(1+\epsilon) \left(\frac{4\pi \mu^2}{M_{\gl}^2}\right)^\epsilon \left\{
-\frac{1}{4\epsilon} - \frac{3}{8} \right\}
\end{equation}
This result can be used to construct the renormalization group equation
for the quark Yukawa coupling without the gluino contribution. The full
renormalization group equation for the running $\overline{MS}$ coupling
including gluon and gluino contributions is at leading order given by
\cite{rgemssm}
\begin{equation}
\mu_R^2 \frac{\partial \bar\lambda_Q(\mu_R)}{\partial \mu_R^2} =
-\frac{C_F}{2} \frac{\alpha_s(\mu_R)}{\pi} \bar \lambda_Q(\mu_R)
\label{eq:yukrge}
\end{equation}
It describes the scale dependence of the running $\overline{MS}$
coupling for scales larger than the quark, squark and gluino masses.
However, if the gluino mass is large compared to the chosen
renormalization scale, the gluino has to be decoupled from the
renormalization group equation. This can be performed consistently by a
momentum subtraction of the gluino contribution for vanishing momentum
transfer while treating the quark and gluon contributions in the usual
$\overline{MS}$ scheme \cite{colwilzee}. It relates the
momentum-subtracted Yukawa coupling $\bar \lambda_{Q,MO}$ to the full
$\overline{MS}$ coupling $\bar\lambda_Q$ in the following way
\begin{equation}
\bar \lambda_{Q,MO}(\mu_R) = \bar\lambda_Q(\mu_R) \left\{ 1 +
C_F\frac{\alpha_s}{\pi} \left(\frac{1}{4}\log\frac{M_{\gl}^2}{\mu_R^2} -
\frac{3}{8} \right) \right\}
\label{eq:yukmoms}
\end{equation}
Differentiating this relation with respect to $\mu_R^2$ yields the
renormalization group evolution of the momentum-subtracted coupling
corresponding to the low-energy result {\it without} the gluino,
\begin{equation}
\mu_R^2 \frac{\partial \bar \lambda_{Q,MO}(\mu_R)}{\partial \mu_R^2} =
-\frac{3}{4} C_F \frac{\alpha_s(\mu_R)}{\pi} \bar \lambda_{Q,MO}(\mu_R)
\label{eq:yukrgemo}
\end{equation}
Since the matching of the effective theory {\it below} the gluino mass
scale to the full MSSM will be performed at $\mu_R=M_{\tilde g}$,
Eq.~(\ref{eq:yukmoms}) determines the required threshold contribution,
too,
\begin{equation}
\bar \lambda_{Q,MO}(M_{\tilde g}) = \bar\lambda_Q(M_{\tilde g}) \left\{ 1 -
\frac{3}{8} C_F\frac{\alpha_s(M_{\gl})}{\pi} \right\}
\label{eq:yukthr}
\end{equation}
In Ref.~\cite{gghnlosqcd} the quark Yukawa coupling has been
renormalized by introducing the quark pole mass $m_Q$. In the effective
theory {\it below} the gluino mass scale the running $\overline{MS}$
coupling is related to the quark pole mass as \cite{msbonsh}
\begin{equation}
g_Q^\phi \frac{m_Q}{v} = \bar \lambda_{Q,MO}(m_Q) \left\{
1+C_F\frac{\alpha_s(m_Q)}{\pi}\right\}
\label{eq:msbonsh}
\end{equation}

\subsection{\boldmath{$\phi\to \tilde Q\bar{\tilde Q}$}}
The analogous calculation has to be repeated for the Higgs coupling to
squarks $\lambda_{\tilde Q}$. The corresponding gluino contribution is
shown in Fig.~\ref{fg:dia}d. In this case no anomalous counter terms are
needed, too. The result of the vertex correction in the heavy gluino
mass limit is given by
\begin{equation}
\tilde Z_1-1 \to C_F \frac{\alpha_s}{\pi} \Gamma(1+\epsilon)
\left(\frac{4\pi \mu^2}{M_{\gl}^2}\right)^\epsilon \left\{
\frac{1}{\epsilon} + 1 \right\}
\end{equation}
Here we concentrate only on the diagonal terms, i.e. the ${\cal H}\tilde
q_L\bar{\tilde q}_L$ and ${\cal H}\tilde q_R\bar{\tilde q}_R$ couplings,
in the no-mixing case since these will be treated by the renormalization
of the Yukawa coupling. In the general case including squark mixing
effects the additional diagonal and non-diagonal contributions will be
absorbed by the renormalized trilinear coupling $A_Q$. The gluino
contribution to the squark wave function renormalization constant $Z_2$
for large gluino masses can be derived from the derivative of the
corresponding self-energy diagram as depicted in Fig.~\ref{fg:self}b,
\begin{equation}
\tilde Z_2-1 \to C_F \frac{\alpha_s}{\pi} \Gamma(1+\epsilon)
\left(\frac{4\pi \mu^2}{M_{\gl}^2}\right)^\epsilon \left\{
-\frac{1}{2\epsilon} - \frac{1}{4} \right\}
\end{equation}
In the case of squark mixing the additional non-diagonal contributions
arising from the self-energy diagram of Fig.~\ref{fg:self}b will be
absorbed by the renormalized mixing angle $\theta_{\tilde q}$.  As for
the quark Yukawa coupling the gluino contribution to the Higgs coupling
to squarks can now be determined,
\begin{equation}
\frac{\Delta \lambda_{\tilde Q}}{\lambda_{\tilde Q}} = \tilde Z_1\tilde
Z_2-1 \to C_F \frac{\alpha_s}{\pi} \Gamma(1+\epsilon) \left(\frac{4\pi
\mu^2}{M_{\gl}^2}\right)^\epsilon \left\{ \frac{1}{2\epsilon} +
\frac{3}{4} \right\}
\end{equation}
For the running $\overline{MS}$ Higgs coupling to squarks the full
renormalization group equation including gluon, squark and gluino
contributions can be expressed as \cite{rgemssm}
\begin{equation}
\mu_R^2 \frac{\partial \bar\lambda_{\tilde Q}(\mu_R)}{\partial \mu_R^2} =
-C_F \frac{\alpha_s(\mu_R)}{\pi} \bar \lambda_{\tilde Q}(\mu_R)
\label{eq:hssrge}
\end{equation}
Note that by virtue of supersymmetry the beta function is twice as large
as the corresponding one for the quark Yukawa coupling in
Eq.~(\ref{eq:yukrge}).  Eq.~(\ref{eq:hssrge}) describes the evolution
for scales above the quark, squark and gluino masses.  The
momentum-subtracted coupling $\bar \lambda_{\tilde Q,MO}$ of the effective
theory without gluinos is now related to the full $\overline{MS}$
coupling $\bar \lambda_{\tilde Q}$ by
\begin{equation}
\bar \lambda_{\tilde Q,MO}(\mu_R) = \bar\lambda_{\tilde Q}(\mu_R) \left\{ 1 -
C_F\frac{\alpha_s}{\pi} \left(\frac{1}{2}\log\frac{M_{\gl}^2}{\mu_R^2} -
\frac{3}{4} \right) \right\}
\label{eq:sqmoms}
\end{equation}
where the additional contributions of quarks, gluons and squarks to
$\bar \lambda_{Q,MO}$ are still $\overline{MS}$ subtracted.  The
evolution of the momentum-subtracted coupling is fixed by the
renormalization group equation, which can be derived by differentiating
Eq.~(\ref{eq:sqmoms}) with respect to $\mu_R^2$,
\begin{equation}
\mu_R^2 \frac{\partial \bar \lambda_{\tilde Q,MO}(\mu_R)}{\partial \mu_R^2} =
-\frac{C_F}{2} \frac{\alpha_s(\mu_R)}{\pi} \bar \lambda_{\tilde Q,MO}(\mu_R)
\label{eq:hssrgeeff}
\end{equation}
This renormalization group equation differs from the corresponding
renormalization group equation of the squared momentum-subtracted Yukawa
coupling $\bar \lambda_{Q,MO}$ of Eq.~(\ref{eq:yukrgemo}). Thus the
supersymmetric relation between these two couplings is violated for
scales {\it below} the gluino mass where the gluino contribution has to
be deleted from the corresponding beta functions. This difference is
expected, since for heavy decoupled gluinos the residual contributing
particle spectrum is not supersymmetric any more. Moreover, for the
matching scale $\mu_R=M_{\tilde g}$, Eq.(\ref{eq:sqmoms}) determines the
threshold correction for the evolution {\it below} the gluino mass scale,
\begin{equation}
\bar \lambda_{\tilde Q,MO}(M_{\gl}) = \bar\lambda_{\tilde Q}(M_{\gl})
\left\{ 1 + \frac{3}{4} C_F\frac{\alpha_s(M_{\gl})}{\pi} \right\}
\label{eq:sqthr}
\end{equation}
Due to the decoupling of the gluinos the soft supersymmetry breaking
induces a hard supersymmetry breaking at low energy scales
\cite{softhard} as can be inferred from the different threshold
corrections and the different renormalization group equations {\it
below} the gluino mass scale.

\subsection{Decoupling of gluinos}
Now we are in the position to derive the effective low-energy scalar
Higgs coupling to gluons. For the consistent decoupling of heavy gluinos
their contribution has to be treated in the momentum-subtraction scheme
as described before. The relation between the momentum-subtracted quark
Yukawa coupling and the scalar Higgs coupling to squarks can be
determined as follows. In the supersymmetric theory, i.e.~for scales
{\it above} the gluino mass, the supersymmetric relation holds for the
running $\overline{MS}$ couplings,
\begin{equation}
\kappa \bar\lambda^2_Q(\mu_R) = \bar\lambda_{\tilde Q}(\mu_R)
\end{equation}
Using Eqs.~(\ref{eq:yukmoms}, \ref{eq:sqmoms}) this leads to a
non-supersymmetric relation between the momentum-subtracted couplings,
\begin{equation}
\kappa \bar \lambda^2_{Q,MO}(m_Q) = \bar \lambda_{\tilde Q,MO}(m_Q) \left\{ 1 +
C_F\frac{\alpha_s}{\pi} \left(\log\frac{M_{\gl}^2}{m_Q^2} - \frac{3}{2}
\right) \right\}
\end{equation}
In this equation we have set the renormalization scale equal to the
quark mass, since at this scale the momentum-subtracted Yukawa coupling
$\bar\lambda_{Q,MO}$ is related to the quark pole mass.  Using
Eq.~(\ref{eq:msbonsh}) for the relation between the quark pole mass and
the $\overline{MS}$ Yukawa coupling of the effective theory below the
gluino mass scale we arrive at the relation
\begin{equation}
2g_Q^{\cal H} \frac{m_Q^2}{v} = \bar \lambda_{\tilde Q,MO}(m_Q) \left\{
1+C_F\frac{\alpha_s}{\pi} \left(\log\frac{M_{\gl}^2}{m_Q^2} +
\frac{1}{2} \right) \right\}
\end{equation}
The proper scale choice for the effective Higgs coupling to squarks,
however, is the squark mass. This choice is relevant for an additional
large gap between the quark and squark masses. Using the renormalization
group equation of Eq.~(\ref{eq:hssrgeeff}) we end up with the final
relation
\begin{equation}
2 g_Q^{\cal H} \frac{m_Q^2}{v} = \bar \lambda_{\tilde Q,MO}(m_{\tilde Q}) \left\{ 1 +
C_F\frac{\alpha_s}{\pi} \left(\log\frac{M_{\gl}^2}{m_{\tilde Q}^2} +
\frac{3}{2}\log\frac{m_{\tilde Q}^2}{m_Q^2} + \frac{1}{2} \right) \right\}
\label{eq:relation}
\end{equation}
The radiative corrections to the relation between the effective
couplings after decoupling the gluinos modify the final result of
Ref.~\cite{gghnlosqcd}. This modification can be discussed in terms of
the effective Lagrangian in the limit of heavy squarks and quarks,
\begin{equation}
{\cal L}_{eff} = \frac{\alpha_s}{12\pi} G^{a\mu\nu} G^a_{\mu\nu}
\frac{\cal H}{v} \left\{\sum_Q g_Q^{\cal H}
\left[1+\frac{11}{4}\frac{\alpha_s}{\pi}\right] + \sum_{\tilde Q}
\frac{g_{\tilde Q}^{\cal H}}{4} \left[1 + C_{SQCD}
\frac{\alpha_s}{\pi}\right] + {\cal O}(\alpha_s^2) \right\}
\end{equation}
where $g_{\tilde Q}^{\cal H}=v\bar \lambda_{\tilde Q,MO}(m_{\tilde Q}) /
m_{\tilde Q}^2$. In Ref.~\cite{gghnlosqcd} the leading term of the
supersymmetric coefficient for large gluino masses has been derived for
equal squark masses $m_{\tilde Q}=M_{\tilde Q_L}=M_{\tilde Q_R}$ as
\begin{equation}
C^{HS}_{SQCD} =
\frac{11}{2} - \frac{4}{3}\log\frac{M_{\gl}^2}{m_{\tilde Q}^2} -
2\log\frac{m_{\tilde Q}^2}{m_{Q}^2}
\end{equation}
where the mismatch between the couplings of Eq.~(\ref{eq:relation}) for
scales below the gluino mass has not been taken into account,
i.e.~keeping the relation $g_{\tilde Q}^{\cal H}=2g_Q^{\cal H} m_Q^2 /
m_{\tilde Q}^2$ after renormalization as in the supersymmetric limit.
Expressing $g_{\tilde Q}^{\cal H}$ in terms of $\bar \lambda_{\tilde
Q,MO}(m_{\tilde Q})$ instead, this mismatch leads to the additional
contribution
\begin{equation}
\Delta C_{SQCD} =
\frac{4}{3}\log\frac{M_{\gl}^2}{m_{\tilde Q}^2} + 2\log\frac{m_{\tilde
Q}^2}{m_{Q}^2} + \frac{2}{3}
\end{equation}
Adding both contributions $C_{SQCD} = C^{HS}_{SQCD}+\Delta C_{SQCD}$
determines the supersymmetric contribution to the effective Lagrangian,
\begin{equation}
C_{SQCD} = \frac{37}{6}
\end{equation}
The resulting effective Lagrangian is well-defined in the limit of large
gluino masses and thus fulfills the constraints of the
Appelquist-Carazzone decoupling theorem \cite{appcar}.

The resulting effective Higgs coupling $\bar \lambda_{\tilde
Q,MO}(m_{\tilde Q})$ can be determined by solving the renormalization
group equations Eqs.~(\ref{eq:yukrgemo}, \ref{eq:hssrgeeff}) which are
valid for scales {\it below} the gluino mass. Taking into account the
proper matching to the fully supersymmetric $\overline{MS}$ coupling at
the scale $\mu_R=M_{\tilde g}$ according to
Eqs.~(\ref{eq:yukthr}, \ref{eq:sqthr}) we arrive at the expression
\begin{equation}
\bar \lambda_{\tilde Q,MO}(m_{\tilde Q}) = 2 g_Q^{\cal H} \frac{m_Q^2}{v}
\frac{1+\frac{3}{2} C_F\frac{\alpha_s(M_{\tilde g})}{\pi}}{1+2
C_F\frac{\alpha_s(m_Q)}{\pi}} \left(\frac{\alpha_s(M_{\tilde
g})}{\alpha_s(m_{\tilde Q})}\right)^{\frac{C_F}{\beta_0}}
\left(\frac{\alpha_s(m_{\tilde
Q})}{\alpha_s(m_Q)}\right)^{\frac{3C_F}{2\beta_0}}
\label{eq:matching}
\end{equation}
where $\beta_0=(33-2N_F-N_{\tilde F})/12$ denotes the leading order beta
function of the strong coupling $\alpha_s$. $N_F$ is the number of
contributing quarks and $N_{\tilde F}$ the number of contributing squark
flavours. Decoupling only the gluino from the supersymmetric spectrum
[$N_F=N_{\tilde F}=6$] its value is given by $\beta_0=5/4$. The
expression (\ref{eq:matching}) can be used to evaluate the effective
coupling $\bar \lambda_{\tilde Q,MO}(m_{\tilde Q})$ from the quark pole
mass to leading logarithmic accuracy. This expression resums the
leading logarithms of the gluino mass and provides the consistent
matching of the low-energy Lagrangian to the full MSSM. In this way the
value of the gluino mass is still measurable as the scale at which the
two couplings $\bar \lambda_{\tilde Q,MO}$ and $\kappa \bar
\lambda^2_{Q,MO}$ merge after taking into account the corresponding
threshold corrections. Note that the pure perturbative expansion of
Eq.~(\ref{eq:matching}) reproduces Eq.~(\ref{eq:relation}) up to ${\cal
O}(\alpha_s)$. Finally we would like to emphasize that
Eqs.~(\ref{eq:relation}, \ref{eq:matching}) do also hold in the general
mixing case for the part of the Higgs couplings to squarks which is
related to the quark Yukawa coupling, if the individual squark masses
are chosen as the renormalization scales.

Since pseudoscalar Higgs bosons only couple to different squark states,
i.e.~to $\tilde Q_L \bar{\tilde Q}_R$ and vice versa, there are no
squark loops at LO so that the pseudoscalar coupling to squarks
contributes at NLO for the first time.  This explains, why the
logarithmic singularity for large gluino masses does not appear in the
pseudoscalar case at NLO \cite{gganlosqcd}.

\section{Conclusions}
In this work we have described the consistent decoupling of heavy gluino
mass effects from the effective Lagrangian for the scalar Higgs
couplings to gluons within the MSSM. We have shown that a careful
extraction of the gluino contributions in the large gluino mass limit
induces a modification of the supersymmetric relations between the Higgs
couplings to quarks and squarks. This modification involves non-trivial
logarithmic gluino mass contributions to the effective couplings. They
exactly cancel the left over gluino mass logarithms of the previous work
of Ref.~\cite{gghnlosqcd} which did not take into account the mismatch
between the Higgs couplings to quarks and squarks at scales below the
gluino mass. This work ensures that the gluino contributions decouple
explicitly for large gluino masses in accordance with the
Appelquist-Carazzone theorem \cite{appcar}. The gluino mass remains in
the effective low-energy theory as the matching scale to the full MSSM.
Analogous methods have to be applied to all processes if decoupling
properties are analyzed. \\

\noindent
{\bf Acknowledgements.} We are indepted to P.M.~Zerwas for reading and
valuable comments on the manuscript. This work is supported in part by
the European Community's Marie-Curie Research Training Network HEPTOOLS
under contract MRTN-CT-2006-035505.

\end{document}